\documentclass [12pt]{article}
\usepackage {graphics}

\begin{document}

\title {Viscous Flows and Conditions for Existence of Shocks in Relativistic Magnetic Hydrodynamics}

\author {V.I.Zhdanov, P.V. Tytarenko, M.S.Borshch \\
\begin{small} \it
Astronomical Observatory of Kiev National Taras Shevchenko
University \end{small}}

\date{}

\maketitle

\begin{abstract}
We present a criterion for a shock wave existence in relativistic
magnetic hydrodynamics with an arbitrary (possibly non-convex)
equation of state. The criterion has the form of algebraic
inequality that involves equation of state of the fluid; it
singles out the physical solutions and it can be easily checked
for any discontinuity satisfying concervation laws. The method of
proof uses introduction of small viscosity into the coupled set of
equations of motion of ideal relativistic fluid with infinite
conductivity and Maxwell equations.
\end{abstract}

\section{Introduction}

It is well known that not every discontinuous solution that can be formally
obtained from equations of hydrodynamics is physically allowed. The most
well known example is prohibition of rarefaction shock waves in media with a
convex equation of state as ones through which entropy is decreasing. But
from the point of view of hydrodynamic equations only, such shock waves have
the same ``rights'' as usual compression shock waves. This example says that
for a shock wave to be physical some additional (non-hydrodynamic)
restrictions must be applied.

In case of usual hydrodynamics (without magnetic field) there are three
famous types of additional criteria that all should hold.

\begin{figure}
\includegraphics[40 mm, 20 mm]{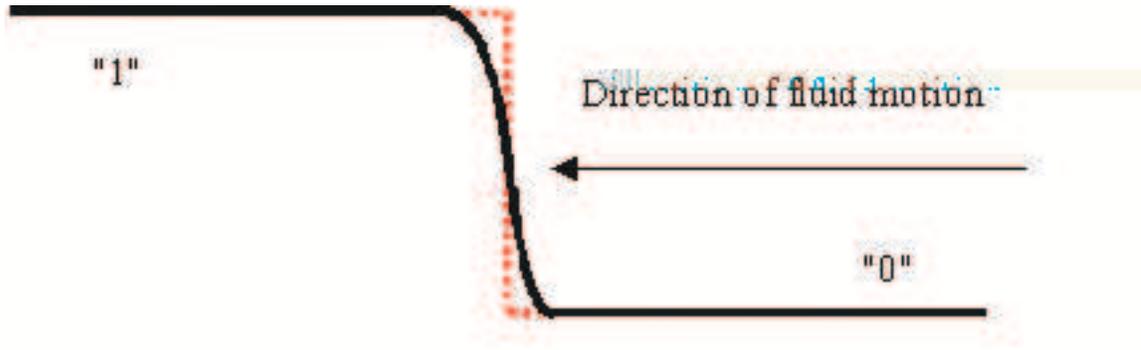}
\caption{Shock wave viscous profile. Dotted line: limiting case of
zero viscosity}
\label{fig1}
\end{figure}

\begin{itemize}
\item Entropy condition. This criterion means that entropy of a
fluid must increase as it is crossing a shock front.

\item Evolutionarity conditions: $v_{ahead} > c_{s} , \quad
v_{behind} < c_{s} $. This criterion follows from condition that
small perturbations must be uniquely defined \cite{landau}.

\item Existence of viscous profile. This condition takes into
account that shock wave is not a sharp step but, due to
dissipative processes, some continuously smeared out region that
we call a profile or structure (see Fig.\ref{fig1}). From
mathematical viewpoint the shock is a generalized solution that
may be represented as a certain limit of viscous flows in zero
viscosity limit \cite{rozh}.
\end{itemize}

If all these conditions hold then one can be sure that a shock
wave exists. In case of "normal" media these criteria are
equivalent. Conditions for normal media are regulated by the set
of Bethe-Weyl conditions \cite{rozh}; and among them the most
important is the requrement of convexity of the equation of state
(EOS), i.e. convexity of the Poisson adiabats.

For a normal EOS all compression shock waves exist and all
rarefaction shock waves are prohibited. However, in case of a
non-convex (anomalous) EOS the question of existence becomes much
more complicated and the criteria 1-3 are not equivalent.
Rarefaction shocks, split shocks, compression simple waves and
complex wave configurations become possible. H.Bethe and H.Weyl
have first studied this situation for shock waves in arbitrary
fluids in classical hydrodynamics (see, e.g.,\cite{rozh}). How the
conditions 1-3 are ordered relatively to each other is discussed
in \cite{men}; the relativistic case is considered in
\cite{tyt_zh}. Note that criterion 3 appears to be the most
effective one and it imposes the most stringent requirements in
case of a general EOS.

In relativistic hydrodynamics the anomalous equation of state
arises in super-dense media particularly with phase transitions;
this has applications in elementary particle physics (see, e.g.,
\cite{bug1, bug2}) and in relativistic astrophysics (super-dense
matter in neutron and exotic stars, gravitational collapse and
supernova explosions, models of gamma ray bursts). The
relativistic version of the convexity condition is

\begin{equation}
{\left. {{\frac{{\partial ^{2}p}}{{\partial \tau ^{2}}}}} \right|}_{S} > 0,
\quad
\tau = \left( {\varepsilon + p} \right) / n^{2}
\end{equation}

where $n$ is the baryon number density, $p$ is the pressure;
$\varepsilon $ is the energy density, $S$ is the entropy per
baryon. In a neighbourhood of phase transitions in super-dense
matter the convexity condition can be violated. The conditions for
existence of relativistic hydrodynamic shocks viscous profile in
case of a general EOS have been investigated in \cite{bug1, bug2}.

The situation in magnetohydrodynamics is more complicated, because
of additional degrees of freedom due to magnetic field. In present
paper we review the main points of investigation of existence
criteria for magnetohydrodynamic shocks. We introduce a viscosity
term into equations of relativistic magnetohydrodynamics in order
to study shock structure and determine conditions for a viscous
profile existence. We derive these conditions for arbitrary EOS on
basis of Landau-Lifshits viscosity tensor with one of viscosity
coefficients put equal to zero. We show that there is a domain of
avoidance that makes our conditions more stringent than
evolutionarity conditions. Then we extend this analysis to both
non-zero viscosity coefficients. Our consideration relies upon the
results of \cite{zh_tyt1}-\cite{zh_bor}.

\section{Basic Equations}

We consider ideal relativistic fluid with infinite conductivity in magnetic
field with energy-momentum tensor

\begin{equation}
\label{eq1}
T^{\mu \nu}  = (p^{\ast}  + \varepsilon ^{\ast} )u^{\mu} u^{\nu}  - p^{\ast
}g^{\mu \nu}  - {\frac{{\mu}} {{4\pi}} }h^{\mu} h^{\nu}  \quad ,
\end{equation}

where $u^{\mu} $ -- four velocity, $F_{\mu \nu} $
--electromagnetic field tensor, $g^{\mu \nu} =g_{\mu \nu}
$=diag(1,--1,--1,--1), $h^{\mu}  = - {\textstyle{{1} \over
{2}}}e^{\mu \alpha \beta \gamma} F_{\alpha \beta} u_{\gamma}  $
(4- vector of magnetic field), $e^{\alpha \beta \gamma \delta} $
-- absolutely anti-symmetric symbol, $\displaystyle p^{\ast}  = p
+ {\textstyle{{\mu}  \over {8\pi}} }{\left| {h} \right|}^{2},
\quad \varepsilon ^{ *}  = \varepsilon + {\textstyle{{\mu}  \over
{8\pi}} }{\left| {h} \right|}^{2}, \quad {\left| {h} \right|}^{2}
= - h^{\alpha} h_{\alpha}>0$, $\mu = $ const -- magnetic
permeability. The tensor (\ref{eq1}) is the sum of corresponding
tensors of hydrodynamics and electrodynamics.

Then we introduce the Landau-Lifshits viscosity tensor
\cite{landau}:
\begin{equation}
\tau _{\mu \nu}  = \eta (u_{\mu ,\nu}  + u_{\nu ,\mu}  - u_{\mu}
u^{\alpha }u_{\nu ,\alpha}  - u_{\nu}  u^{\alpha} u_{\mu ,\alpha}
) + (\xi - 2\eta / 3)u^{\alpha} _{,\alpha}  (g_{\mu \nu}  -
u_{\mu}  u_{\nu}  ),
\end{equation}
into equations of motion
\begin{equation}
\label{eq2}
\partial _{\mu}  (T^{\mu \nu}  + \tau ^{\mu \nu} ) = 0,
\end{equation}

\begin{equation}
\label{eq3}
\partial _{\mu}  (nu^{\mu} ) = 0,
\end{equation}

\begin{equation}
\label{eq4}
\partial _{\mu}  (u^{\mu} h^{\nu}  - u^{\nu} h^{\mu} ) = 0 \quad .
\end{equation}

Here Eq. (\ref{eq2}) describes energy-momentum conservation, Eq.
(\ref{eq3}) -- baryon number conservation; Eq.(\ref{eq4}) follows
from Maxwell equations for the electromagnetic field.

In order to obtain shock structure in the rest frame of the shock we have to
take a one-dimensional stationary solution of (\ref{eq2})--(\ref{eq4}) depending upon the
only variable $x$ corresponding to the direction perpendicular to the shock
front. This yields
\begin{equation}
\label{eq5}
T^{1\nu}  + \tau ^{1\nu}  = T_{(0)}^{1\nu}  ,
\end{equation}

\begin{equation}
\label{eq6}
u^{1}h^{\nu}  - h^{1}u^{\nu}  = H^{\nu}  \equiv u_{(0)}^{1} h_{(0)}^{\nu}  -
h_{(0)}^{1} u_{(0)}^{\nu}
\end{equation}

\begin{equation}
\label{eq7}
nu^{1} = n_{(0)} u_{(0)}^{1}
\end{equation}

If we denote the state ahead of the shock transition by index "0"
then relation between states on both sides of the shock (with
state "1" being the state behind a shock) must be as follows
\begin{equation}
\label{eq8}
T_{(\ref{eq1})}^{1\nu}  = T_{(0)}^{1\nu}  ,
\end{equation}

\begin{equation}
\label{eq9}
u_{(\ref{eq1})}^{1} h_{(\ref{eq1})}^{\nu}  - h_{(\ref{eq1})}^{1} u_{(\ref{eq1})}^{\nu}  = H^{\nu}  \equiv
u_{(0)}^{1} h_{(0)}^{\nu}  - h_{(0)}^{1} u_{(0)}^{\nu}  ,
\end{equation}

\begin{equation}
\label{eq10}
n_{(\ref{eq1})} u_{(\ref{eq1})}^{1} = n_{(0)} u_{(0)}^{1} \quad .
\end{equation}

By definition we say that shock wave "0"$ \to $"1" has a viscous
profile if there is a continuous solution of
(\ref{eq5})--(\ref{eq7}) having corresponding asymptotics for
\textit{x$ \to $}$\infty $ and \textit{x$ \to $} --$\infty $.

\section{The case of one viscosity coefficient $\eta $=0.}

One can choose the reference frame such as $u^{3} \equiv 0$ and $h^{3}$.
Then one can obtain
\begin{equation}
\label{eq11}
\varepsilon ^{\ast} u^{1} = T_{(0)}^{1\mu}  u_{\mu}
\end{equation}
from (\ref{eq5}) and
\begin{equation}
\label{eq12} h^{\mu}  = {\frac{{1}}{{u^{1}}}}{\left[ {H^{\mu}  -
u^{\mu} H^{\alpha }u_{\alpha}}   \right]}
\end{equation}
from (\ref{eq6}).

From continuity equation
\begin{equation}
\label{eq13}
n = u_{\left( {0} \right)}^{1} n_{0} / u^{1},
\end{equation}
then Eqs. (\ref{eq11})-(\ref{eq14}) allow us to express all the
variables in terms of $u^{1}$ and $u^{2}$.

Using this one can get then relationship between $u^{1}$ and
$u^{2}$
\begin{equation}
\label{eq14}
\left( {H^{2}u^{2} - H^{0}u^{0}} \right)\left( {H^{2}u^{0} - H^{0}u^{2}}
\right) = 4\pi u^{1}\left( {T_{{\rm (}{\rm 0}{\rm )}}^{10} u^{2} - T_{{\rm
(}{\rm 0}{\rm )}}^{12} u^{0}} \right),
\end{equation}

This allows us to relate three-dimensional velocity components
$v_{1}=u^{1}$/$u^{0}$ and $v_{2}=u^{2}$/$u^{0}$ with each other.
Note that we do not consider here the trivial case, when the
magnetic field is zero. We have an explicit dependence
$v_{1}(v_{2})$ from (\ref{eq14}) that leaves us the only ordinary
differential equation for $u^{1}$
\begin{equation}
\label{eq15}
\zeta \left( {1 + \left( {u^{1}} \right)^{2}} \right){\frac{{du^{1}}}{{dx}}}
= T^{11} - T_{{\rm (}{\rm 0}{\rm )}}^{11} ,
\end{equation}
that describes the viscous structure of profile. Our problem is
reduced to find a necessary condition for a regular solution of
Eq.(\ref{eq15}) to exist. We assume also that hydrodynamic
parameters in the shock structure are monotonous functions. If we
suppose that such solution exists then we have finite
\textit{dv}$_{1}$/\textit{dv}$_{2}$ and for monotonous structures
this cannot change its sign. Then the function $v_{2}(v_{1})$ is
well defined during the transition "0"$ \to $"1". In this case the
right hand side of (\ref{eq15}) must not change its sign;
otherwise the hydrodynamic parameters could not reach the point
"1". This yields necessary condition we are looking for. For
details see \cite{zh_tyt1, zh_tyt2}.

In view of (\ref{eq7}) we can express $v_{1}$ and $v_{2}$ in terms
of the specific volume $V=$1/$n.$

If introduce the function
\[
\begin{array}{l}
  \tilde {p}\left( {V} \right) = \{1 + {\rm (}u^{1}{\rm )}^{2}{\rm
\}}^{ - 1}{\rm \{}T_{{\rm (}{\rm 0}{\rm )}}^{{\rm 1}{\rm 1}} +
{\frac{{1}}{{4\pi}} }{\rm (}H^{\alpha} u_{\alpha}  {\rm )}^{2} - T_{{\rm
(}{\rm 0}{\rm )}}^{1\mu}  u_{\mu}  u^{1}{\rm \}} - \\
 - {\frac{{1}}{{8\pi}} }{\rm (}u^{1}{\rm )}^{ - 2}{\rm \{}{\rm (}H^{\alpha
}u_{\alpha}  {\rm )}^{2} - H^{\alpha} H_{\alpha}  {\rm \}} \\
 \end{array};
\]
where $\tilde {\varepsilon} {\rm (}V{\rm )} = T_{{\rm (}{\rm
0}{\rm )}}^{1\mu}  u_{\mu}  / u^{1} - {\textstyle{{1} \over
{8\pi}} }{\left| {h} \right|}^{2}$, then we obtain following
\textit{criterion of admissibility of stationary shock
transition}:

\begin{equation}
\label{eq16}
\left( {V_{1} - V_{0}}  \right)\left( {p\left( {V,\tilde {\varepsilon
}\left( {V} \right)} \right) - \tilde {p}\left( {V} \right)} \right) \ge 0
\end{equation}
for all $V$ between $V_{0}$ and $V_{1}$.

This criterion has the following consequences [7,8]:

\begin{itemize}
\item If the shock satisfies the criterion, then it also satisfies
the entropy criterion.

\item The criterion can be written in terms of the shock adiabat
$p_{H} \left( {V} \right)$ as follows
\begin{equation}
\left( {V_{1} - V_{0}}  \right)\left( {p_{H}
\left( {V} \right) - \tilde {p}\left( {V} \right)} \right) \ge 0
\quad .
\end{equation}

In case of a non-single valued shock adiabat the criterion can be
formulated in terms of absence of intersections of $p_{H} \left(
{V} \right)$ and $\tilde {p}\left( {V} \right)$ everywhere except
$V_{0} $ and $V_{1} $.

\item In the neighbourhoods of initial and final points we get that the
possible \textit{fast shock} transitions are found to satisfy the following relations between
the speed of the shock and the characteristic speeds at initial state ``0''
(ahead of the shock)
\begin{equation}
v_{sh(0)} > v_{f(0)} > v{}_{N(0)} > v_{A(0)} > v_{sl(0)}
\end{equation}
and at the final state ``1'' (behind the shock).
\begin{equation}
v_{f(\ref{eq1})} > v_{sh(\ref{eq1})} > v{}_{N(\ref{eq1})} >
v_{A(\ref{eq1})} > v_{sl(\ref{eq1})}
\end{equation}
\end{itemize}

Here we have introduced a new characteristic velocity that
satisfies equation
\begin{equation}
\left( {p + \varepsilon}  \right){\rm (}u_{N}^{1} {\rm )}^{2}{\rm
[}{\rm (}u_{N}^{1} {\rm )}^{2} + 1{\rm ]} = {\frac{{1}}{{4\pi}}
}{\rm (}h^{1}{\rm )}^{2}
\end{equation}

It is distinct from Alfven velocity that enters usual
evolutionarity conditions.

For \textit{slow shocks} we have the usual evolutionarity conditions.
\begin{equation}
v_{A(0)} > v_{sh(0)} > u_{sl(0)} \quad ; \quad v_{f(\ref{eq1})} >
v{}_{N(\ref{eq1})} > v_{A(\ref{eq1})} > v_{sl(\ref{eq1})} >
v_{sh(\ref{eq1})}.
\end{equation}

So we have obtained a domain of avoidance behind the fast shock
${\left[ {v_{N} ,v_{A}}  \right]}$ that satisfies evolutionarity
conditions but nevertheless must be considered as non-physical one
as according to our criterion there is no shock wave with viscous
profile there. Note that the existence of $v_{N}$ is due to
monotonicity of the dependence $v_{1}(v_{2})$.

This relations for the final state appear to be more restrictive than the
standard evolutionary criterion that does not involve $V_{N}$ . However in
the nonrelativistic limit $V_{N} \to V_{A}$ and these inequalities tend to
the standard evolutionarity conditions.

\section{Two nonzero viscosity coefficients}

In this case when $\eta \ne 0$ instead of algebraic equation
(\ref{eq13}) we have the second ordinary differential equation
(for $u^{2})$. It is convenient to introduce a new variable $v =
{{u^{2}} \mathord{\left/ {\vphantom {{u^{2}} {\sqrt {1 +
(u^{1})^{2}}}} } \right. \kern-\nulldelimiterspace} {\sqrt {1 +
(u^{1})^{2}}}} $.

After this we have the following dynamical system regarding to
$u^{1}$ and $v$
\begin{equation}
\label{eq17}
\begin{array}{l}
 \displaystyle \left( {\xi + 4\eta / 3} \right){\frac{{du^{1}}}{{dx}}} = F_{1}
(u^{1},v),     \\
   \displaystyle \eta {\frac{{dv}}{{dx}}} = F_{2} (u^{1},v) \\
 \end{array},
\end{equation}
where
\[
F_{1} (u^{1},v) = p - (1 + (u^{1})^{2})^{ - 1}{\left[ {T_{0}^{11}
+ {\frac{{\mu}} {{4\pi}} }(H^{\alpha} u_{\alpha}  )^{2} -
T_{0}^{1\mu}  u_{\mu } u^{1}} \right]} +{\frac{{\mu}} {{8\pi}}
}(u^{1})^{ - 2}[(H^{\alpha} u_{\alpha}  )^{2} - H^{\alpha}
H_{\alpha}  ]
\]
\[
F_{2} (u^{1},v) = {\frac{{(T_{0}^{10} u^{0} - T_{0}^{12} u^{2})u^{2}u^{1} -
(\mu / 4\pi )u^{2}(H^{\alpha} u_{\alpha}  )^{2}}}{{4\pi u^{1}[1 +
(u^{1})^{2}]^{{\frac{{5}}{{2}}}}}}} +
\]
\[
 + {\frac{{(\mu / 4\pi )H^{2}(H^{\alpha} u_{\alpha}  ) - T_{0}^{12}
u^{1}}}{{[1 + (u^{1})^{2}]^{3 / 2}u^{1}}}}
\]

Let the state parameters $u_{{\rm (}0{\rm )}}^{\mu}  ,h_{{\rm
(}0{\rm )}}^{\mu}  ,n_{0} ,p_{0} $ ahead of the shock and $u_{{\rm
(}1{\rm )}}^{\mu} ,h_{{\rm (}1{\rm )}}^{\mu}  ,n_{1} ,p_{1} $
behind the shock satisfy the conservation laws
(\ref{eq8})--(\ref{eq10}) that relate hydrodynamic quantities on
both sides of the shock. We denote $y=u^{1}$, $y_{0}=u_{(0)}^{1}
$, $y_{1}=u_{(\ref{eq1})}^{1}$. In this section we deal with the
state variables $u^{\mu} $, $h^{\mu} $, $^{} n$, $ p$ ahead of the
shock unless otherwise stated, and we omit further the index "0"
for these variables.

Now we rewrite the conditions of viscous profile existence in terms of the
right hand sides of the system (\ref{eq17}) and the curves $V_{1}$,$ V_{2}$:

$v=V_{1}(y)$: $F_{1} (y,V_{1} (y)) = 0$

$v=V_{2}(y)$: $F_{2} (y,V_{2} (y)) = 0$

$v_{1}=V_{1}(y_{1})=V_{2}(y_{1})$;
$v_{0}=V_{1}(y_{0})=V_{2}(y_{0})$, $V_{1}(y) \ne V_{2}(y) \quad
\forall y \in (y_{1}$,$y_{0})$

The conditions can be formulated as follows

\begin{description}
\item[A] The function $v=V_{2}(y)$ is monotonous on ($y_{1}$,
$y_{0})$.

\item[B] For $y \in (y_{1}$,$y_{0})$ the following inequality is valid:
\[
(y_{0} - y_{1} )F_{1} (y,V_{2} (y)) < 0 \quad ,
\]

\item[C] We suppose that $h^{1}h^{2} \ne $0 at the point "0" (this
is a technical requirement).

\item[D] (\textbf{D1}): $u^{1}>u_{f}$ ahead of the shock at the point "0"
and $u_{N}<u^{1}<u_{f}$ behind the shock at the point "1", or
(\textbf{D2}): $u_{A}>u^{1}>u_{sl}$ ahead of the shock at the
point "0" and $u_{sl}>u^{1}$ behind the shock at the point "1.
\end{description}

\begin{figure}
\includegraphics[40 mm, 30 mm]{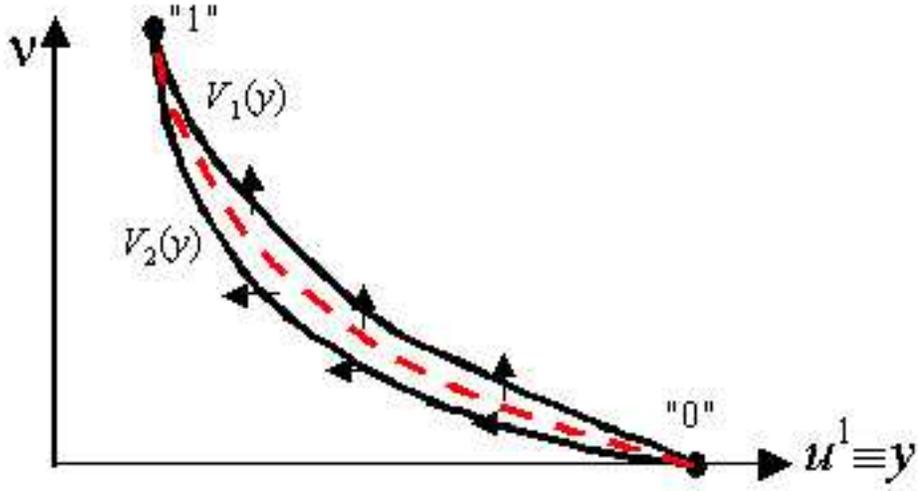}
\caption{Fast shock. The arrows show direction of the phase
trajectories when crossing the curves $V_1$ and $V_2$. Red dashed
line corresponds to viscous profile of the shock.} \label{Fig2}
\end{figure}

The last inequalities are the famous relations between the
velocities ahead of the shock. The first inequality (D1)
corresponds to the fast shock and the second one (D2) -- to the
slow shock. In case $\eta $=0 inequalities (D1) and (D2) follow
from criterion of viscous profile existence of the section
(\ref{eq3}).

Under these conditions one can show, e.g., that in case D1 the rest point
"1" of the system (\ref{eq15}) is a saddle point. In case D2 the rest point "0" is a
saddle point. This enables us to restore the qualitative behaviour of
solutions to the system (\ref{eq17}) inside the domain of the phase plane between
the curves $V_{1}$ and $V_{2}$.

Typical situation is shown on Figs.2 and 3. In case of the fast
shock (D1) the solutions leaving "0" go out of the domain, except
the only solution representing the separatrix of the saddle point
"1"; just this solution represents the shock viscous structure.

In case of the slow shock (D2) there is the only solution leaving
"0" that tends to "1"; just this solution represents the shock
viscous structure. The other solutions enter the domain, crossing
$V_{2}$ from right to left and $V_{1}$ bottom-up.

As the first result we have obtained following sufficiency of
conditions for $\eta>0$, $\xi>0$.

\textit{Let the states ``0'' ahead of the shock and ``1'' behind
the shock satisfy the conservation equations. If the conditions
(A--D) are satisfied, then the MHD shock transition "0"$ \to $"1"
has a viscous profile satisfying the equations
(\ref{eq5})--(\ref{eq10}).}

\begin{figure}
\includegraphics[40 mm, 50 mm]{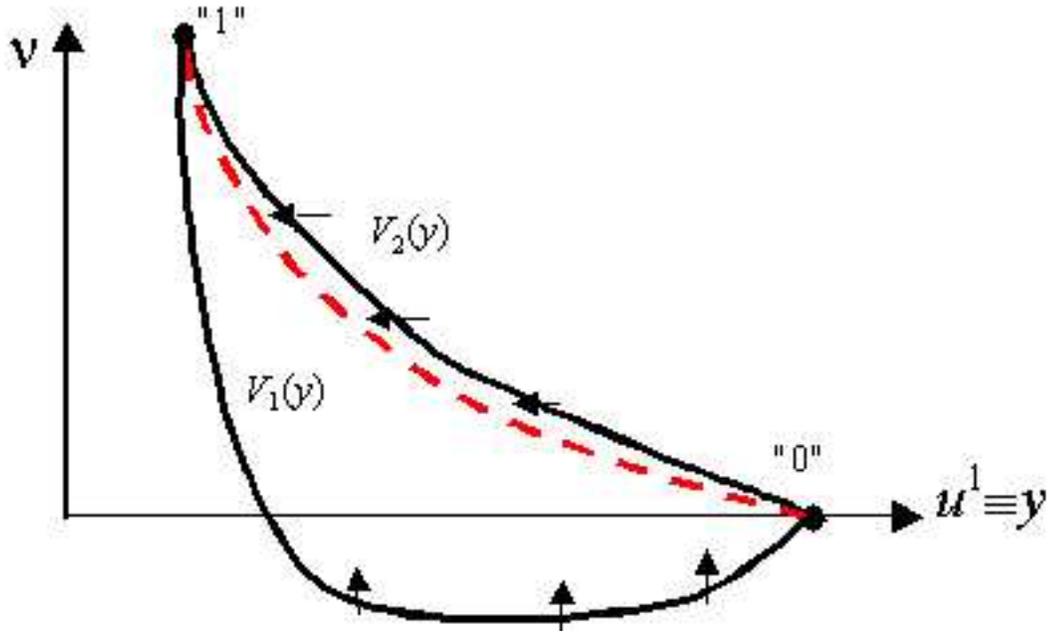}
\caption{Slow shock. The arrows show direction of the phase
trajectories when crossing the curves $V_1$ and $V_2$. Red dashed
line corresponds to viscous profile of the shock.} \label{Fig3}
\end{figure}

More detailed proof will be published elsewhere \cite{zh_bor}.

If additional limitations on EOS (e.g., convexity) are absent, the
criterion (B) dealing with EOS for the whole interval between the
states "0" and "1" is evidently more restrictive than, e.g.,
evolutionarity conditions [9] or any other conditions that involve
characteristics of the fluid only at initial and final states. On
the other hand, the evolutionarity conditions can be derived from
the viscous profile existence \cite{zh_tyt2, zh_bor}.

Our criteria can be applied to an arbitrary smooth EOS. However,
we must note that the requirement for $V_{1}(y)$ to be a
continuous (single-valued) function is not trivial and can not be
fulfilled in case of a certain equations of state (see, e.g.,
\cite{rozh, men}). Though consideration of a viscous profile seems
to be rather effective for investigation of shock existence and
stability, this method can not work in case of complicated EOS
(see, e.g., remarks in [3] in case of relativistic hydrodynamics)
that require either modification of the equations of motion or
using additional physical information about solutions.

\end{document}